\documentclass[a4paper,11pt]{article}
\usepackage{pos}
\usepackage{natbib}
\usepackage{wrapfig}

\title{The contribution of distant sources to the observed flux of ultra high-energy cosmic rays}
 \ShortTitle{Ultra high-energy CRs from distant sources}

\author*[a,b]{Ellis R. Owen}
\author[c,d]{Qin Han}
\author[d,e]{Kinwah Wu}
\author[a]{Y. X. Jane Yap}
\author[f]{Pooja Surajbali}

\affiliation[a]{Institute of Astronomy, National Tsing Hua University, Hsinchu, Taiwan (ROC)}
\affiliation[b]{Center for Informatics and Computation in Astronomy, National Tsing Hua University, \\Hsinchu, Taiwan (ROC)}
\affiliation[c]{School of Astronomy and Space Science, Nanjing University, 163 Xianlin Avenue, Nanjing, \\Jiangsu 210093, People's Republic of China}
\affiliation[d]{Mullard Space Science Laboratory, University College London, Holmbury St. Mary, Dorking, \\Surrey RH5 6NT, United Kingdom}
\affiliation[e]{Research center for Astronomy, Astrophysics and Astrophotonics, Macquarie University, \\Sydney, NSW 2109, Australia}
\affiliation[f]{Max-Planck-Institut f\"{u}r Kernphysik, Saupfercheckweg 1, Heidelberg 69117, Germany}

\emailAdd{erowen@gapp.nthu.edu.tw}

\abstract{Ultra-high-energy (UHE) cosmic rays (CRs) interact  with cosmic background radiation through hadronic processes, and the Universe would become `opaque' to UHE CRs of energies $\sim$($10^{18}$- $10^{20}$) eV over about several tens of Mpc, setting the Greisen-Zatsepin-Kuz'min (GZK) horizon. We demonstrate that a non-negligible fraction of the UHE CRs arriving on Earth could originate from beyond the GZK horizon when heavy nuclear CRs, and the population and evolution of UHE CR sources are taken into account. We show how the multi-particle CR horizon is modified by different source populations, and discuss how this leads to the natural emergence of an isotropic flux component in the observed UHE CR background. This component would coexist with an anisotropic foreground component contributed 
by nearby sources within the GZK horizon.}

\FullConference{37$^{\rm{th}}$ International Cosmic Ray Conference (ICRC 2021)\\
		July 12th -- 23rd, 2021\\
		Online -- Berlin, Germany}

\begin{document}
\maketitle

\section{Introduction}

\noindent
The detection rate of ultra high-energy (UHE) cosmic ray (CR) events on Earth is about 1 particle per square km per century at $E=10^{20}\;\!{\rm eV}$, corresponding to a flux of $E^3 J(E) \approx 10^{24} \;\!{\rm eV}^2\;\!{\rm m}^{-2}\;\!{\rm s}^{-1}\;\!{\rm sr}^{-1}$~\citep{Ivanov2017wH}.   
Several origins of these particles have been proposed, where conditions exist to allow acceleration of particles to 
ultra-high energies over large scales or in strong magnetic fields~\citep[cf. the Hillas criterion, ][]{Hillas1984ARAA}, such as in the relativistic jets of active galactic nuclei (AGN), neutron stars and in the large-scale shocks associated with galaxy clusters. 
However, their exact origins remain unsettled.

The low flux of UHE CRs is attributed both to the power-law nature of the intrinsic emission spectra from plausible sources~\cite{Fermi1949PhRv}, coupled with the attenuation of cosmologically propagating UHE particles with ambient radiation fields, leading to the development of a high-energy `ankle' in the CR spectrum, i.e the Greisen-Zatsepin-Kuz'min (GZK) effect~\cite{Greisen1966PhRvL, Zatsepin1966JETPL}.   
The attenuation of UHE CRs is mainly caused by hadronic interactions with the cosmological microwave background (CMB) and, to a lesser extent, the infrared and optical components of the extra-galactic background light (EBL).   
The Universe becomes `optically thick' to UHE CRs 
over distances of a few 10s Mpc (at energies of $E\approx 10^{20}\;\!{\rm eV}$), setting a GZK `horizon'.  
The extent of this horizon depends on the energy 
and nuclear mass of a CR particle. 
In general, more energetic particles are attenuated 
over shorter distances.  
CRs of higher atomic number would undergo erosive processes 
(e.g. photo-spallation), 
allowing their surviving descendant particles 
to continue propagating over a large distance.   

The extent of the GZK `horizon' 
is therefore not a sharp upper-limit  
on the distance over which UHE CRs are allowed propagate.  
It is instead a rough distance scale over which 
a substantial reduction of the CR flux attenuation would be expected.  
Thus, UHE CRs arriving on Earth  
(1) originate from either the nearby sources within this GZK horizon, and/or (2) are 
the residual attenuated flux from 
populations of sources distributed over cosmological distances. 
If sources are predominantly distant, and the detected UHE CRs on Earth comprises a substantial component of their residual attenuated flux contribution, an isotropic diffuse background of UHE CRs would be expected. Support for this has emerged in observational studies, with correlations between UHE CR arrival directions and possible local sources being strengthened if the UHE CR flux arriving on Earth is assumed to be comprised of a diffuse isotropic \textit{background} component from distant source populations superposed with an anisotropic \textit{foreground} flux originating from nearby cosmic accelerators~\citep{Kim2011JCAP, Aab2017Sci}. 
In this work~\cite[see also][for further details and discussion]{Owen2021arXiv210712607O}, we consider the emergence of such a \textit{background} component arising from distant source populations. 
We investigate four possible source populations (adopting the fitted models of~\cite{Batista2019JCAP}), and show that the emerging UHE CR background can contribute a substantial fraction of the UHE CR flux arriving on Earth.

\section{Interactions of CR nuclei in cosmological radiation fields}
\label{sec:section_2}

\subsection{CR interactions}
\label{sec:section_21}

\noindent
UHE CR nuclei propagating through intergalactic space predominantly interact with ambient radiation fields (1) via Beth-Heitler (BH) photo-pair production~\citep{Bethe1934PRSA}, which we model as a cooling process; (2) Photo-pion production~\citep{Berezinsky1993PRD}, which we model as an absorption process, and (3) Photo-spallation~\citep{Puget1976ApJ}, which we model as a disintegration process for nuclei with mass number $A>1$. Other processes, such as radiative cooling and adiabatic cooling of the cosmologically-propagating particles also operate. 
In a black-body radiation field, the cooling rate due to BH pair-production for a CR of energy $E_A = A\;\!\gamma_{A} m_{\rm p} c^2$ (where $m_{\rm p}$ is the proton rest mass, and $c$ is the speed of light) is
\begin{equation}%
\label{eq:photopair_losses}%
 {b^{\rm \gamma A}}_{\! A}(\epsilon_A) 
\approx 
	- \frac{{f\;\!Z_A^2}\;\!{c} \alpha_{\rm f} \sigma_{\rm T}\mathcal{F}_{\rm \gamma e}(u)}{{\lambda_{\rm c}}^3 {\gamma_{A}}^2{A}\;\! u^{5/2}}   \ ,  
\end{equation}%
where the dimensionless variable $u =  (\gamma_A \Theta)^{-1}$, $\Theta = k_{\rm B}\;\!T/m_{\rm e}c^2$ is the dimensionless temperature ($k_{\rm B}$ is the Boltzmann constant, $m_{\rm e}$ is the electron rest-mass)
 and the function $\mathcal{F}_{\rm \gamma e}(u)$
 takes the same form as that in \citep{Owen2019AA}. $f$ is the dilution factor of the radiation field ($f = 1$ for the CMB), $\alpha_{\rm f}$ is the fine structure constant, $\lambda_{\rm c}$ is the Compton wavelength of an electron, and $\sigma_{\rm T}$ is the Thomson cross section.
 Radiative cooling is dominated by Compton scattering, and arises at a rate of 
\begin{align}%
\label{eq:radiative_losses}%
b_{\rm rad}(\epsilon_A, z) 
  &= - \frac{4}{3}\frac{\sigma_{\rm T}\;\! U(z)}{m_{\rm e} c} 
  \left[
   \frac{{\epsilon_A}^{1/2}Z_A}{A} 
   \left(\frac{m_{\rm e}}{m_{\rm p}}\right)      
   \right]^4 \  
\end{align}%
\cite{Puget1976ApJ},   
where $U(z)$ is the radiation energy density 
(at redshift $z$), $\epsilon_{A} = E_{A}/m_{\rm e} c^2$, and $Z_A$ is the nuclear charge.\footnote{In this work, we relate atomic mass to nuclear charge using the relation presented in~\cite{Puget1976ApJ}.} Adiabatic losses experienced by UHE CR nuclei due to cosmological expansion occur at a rate of $b_{\rm ad}(\epsilon_{A}, z) 
  = - \epsilon_{A} \;\! H_{\rm 0}\;  \mathcal{E}(z)$, 
 for $\mathcal{E}(z)=\left[\Omega_{\rm r, 0}(1+z)^4 + \Omega_{\rm m, 0}(1+z)^3 + \Omega_{\rm \Lambda, 0} \right]^{-1/2}$,  
  where $\Omega_{\rm m,0} = 0.315\pm0.007$, 
  $\Omega_{\rm r,0} \approx 0$ 
  and $\Omega_{\rm \Lambda,0} = 0.685\pm0.007$ 
  are the normalized density parameters for matter, 
  radiation and dark energy respectively.  
The present value of the Hubble parameter 
  $H_0 = 100\;\! h\;\! {\rm km}\;\!{\rm s}^{-1}{\rm Mpc}^{-1}$,
  where $h = 0.673\pm 0.006$ 
  \cite{Planck2018}. 
 
If approximating the photo-pion interaction cross section as a delta function~\cite{Unger2015PhRvD}, the CR interaction rate in a black-body radiation field can be written as 
\begin{align}
\label{eq:rate_pA}
{{R}^{A\pi}}_{\! A} 
&=- \frac{2\pi^2 \;\!{f}\;\! c \sigma_0 \Gamma_{\rm res}A \epsilon_{\rm res} \Theta}{{\gamma_{A}}^2{\lambda_c}^3}\ln\left[1-\exp\left(\frac{-\epsilon_{\rm res}}{2\gamma_{A} \Theta}\right)\right] \ , 
\end{align} 
where $\sigma_0 = 5.0\times 10^{-28}\;\!{\rm cm}^2$, $\Gamma_{\rm res} = 150\;\!{\rm MeV}$, and $\epsilon_{\rm res} \approx 665$. The corresponding particle absorption rate is then ${\Lambda^{A \pi}}_{\! A}={{{R}^{A\pi}}_{\! A}}/{A}$.
 The photo-disintegration rate takes a similar form, if again approximating the interaction cross section with a delta-function~\cite{Wang2008ApJ} and assuming single-nucleon emission:
\begin{align}
\label{eq:approx_photo_R}
 R^{A'}_{A}
     = - \frac{4\pi \;\!{f} \;\! c \sigma_{\rm 0,A} \Delta_n \epsilon_0 \Theta}{{\gamma_A}^2{\lambda_c}^3} 
       \ln\left[\;\!1-\exp\left(\frac{-\epsilon_0}{2\gamma_A \Theta}\right)\right] \ . 
\end{align} 
Here, $\sigma_{0, A} = 1.45\times 10^{-27}\;\! A \;\!{\rm cm}^2$, 
$\Delta_n = 8\;\!{\rm MeV}$, and $\epsilon_0 = 83.46 A^{-0.21}$ for $A>4$, and $1.81 A^{2.433}$ for $A\leq 4$~\cite{Karakula1993APh}.
Strictly, this is equivalent to the production rate of $A'=1$ or $A' = A-1$ nuclei due to the spallation of nucleus of mass number $A$. The corresponding particle absorption (or secondary nuclear production) rate, if assuming single-nucleon emission in each photo-disintegration process, is given by ${\Lambda^{A'}}_{\!A} \approx R^{A'}_{A}/{A}$.  The total absorption rate for UHE CR nuclei is then given by $\Lambda_A={\Lambda^{\rm sp}}_{\! A}+{\Lambda^{A \pi}}_{\! A}$, while the
 source term for secondary nuclei produced in photo-spallation follows as:
\begin{align}
{Q^{\rm sp}}_{\! A}(z, \epsilon_A) &   \approx \frac{1}{(1+z)^3} \begin{cases}
     \int_{\epsilon_{\rm min}}^{\epsilon_{\rm max}} {\rm d}\epsilon_{{A}'} {R^{A'}_{A} n_{{A}'}} \ , \hspace{4.7cm} {\rm if} \; A>1{\rm ;}\\
    2\int_{\epsilon_{\rm min}}^{\epsilon_{\rm max}} 
  {\rm d}\epsilon_{2} {R^{A'}_{A} n_{2}} +  \sum_{{A}'} \int_{\epsilon_{\rm min}}^{\epsilon_{\rm max}} 
  {\rm d}\epsilon_{{A}'} R^{A'}_{A} n_{{A}'} \ , \hspace{1cm} {\rm if} \; A=1{\rm .}
    \end{cases} \ ,
    \label{eq:source_secondary}
\end{align}   
which proceeds while CR energies (including the secondary nuclei) have sufficient energy to continue to interact, thus forming an extended particle cascade along UHE CR propagation paths.

\subsection{Cosmological radiation fields}

\begin{wraptable}{r}{0.5\textwidth}
\centering
\resizebox{0.48\textwidth}{!}{%
\begin{tabular}{lccc}
\hline \hline
{\bf Component} & {\bf $\Theta_{\boldsymbol{i}}(\boldsymbol{z})$}  & {\bf $\boldsymbol{f}_{\boldsymbol{i}}(\boldsymbol{z})$} & \\ \hline
     {\it CMB} & $4.58\times 10^{-10} \;\! (1+z)$ & $1.0$ \\
    {\it EBL -- Dust} & $1.1\times10^{-8}$ &  $3.5\times10^{-7}$\;\!$\mathcal{K}(z)$ \\
    {\it EBL -- UV/O (1)} & $6.7\times 10^{-8}$ & $9.0\times10^{-12}$\;\!$\mathcal{K}(z)$ \\
    {\it EBL -- UV/O (2)} & $1.7\times10^{-7}$ & $4.5\times 10^{-13}$\;\!$\mathcal{K}(z)$ \\
    {\it EBL -- UV/O (3)} & $5.1\times10^{-7}$ &  $1.4\times 10^{-14}$\;\!$\mathcal{K}(z)$ \\
    {\it EBL -- UV/O (4)} & $9.3\times10^{-7}$ &  $6.0\times 10^{-16}$\;\!$\mathcal{K}(z)$ \\
    {\it EBL -- UV/O (5)} & $2.0\times10^{-6}$ &  $6.1\times 10^{-17}$\;\!$\mathcal{K}(z)$ \\
    \hline \hline
\end{tabular}}
\caption{
Summary of parameter choices radiation field components. The function $\mathcal{K}(z)$ is given by equation~\ref{eq:kevo}.}  
\label{tab:radiation_params} 
\end{wraptable}

\noindent
The CMB permeates the Universe and provides a strong target field to drive the interactions described in section~\ref{sec:section_21}. Additionally, the EBL can also act to attenuate more massive UHE CRs with energies slightly lower than those principally affected by the CMBs~\cite{Aloisio2013APh, Owen2021arXiv210712607O}. The EBL is comprised of radiation emitted from astrophysical objects~\citep[for an overview, see][]{Cooray2016RSOS}. Its energy is mainly concentrated in two spectral peaks -- one at optical wavelengths, being broadly associated with stellar emission from within populations of galaxies, while the other is at infrared wavelengths, and is presumably dominated by dust-reprocessed astrophysical emission.

We model the CMB as a black-body radiation field, and adopt a simple analytic representation of the EBL, comprised of five superposed diluted black-body components, given by 
\begin{equation}
   n_i(\epsilon_{\rm ph}, z)
  = \frac{8\pi\;\! f_{i}(z)}{\lambda_{\rm c}^3}
    \frac{\epsilon_{\rm ph}^2}{\exp (\epsilon_{\rm ph}/{\Theta}_i(z))-1} \ ,  \label{eq:general_form_ebl}
\end{equation}
where one component is attributed to dust (infrared) emission and the rest approximate the ultraviolet/optical (UV/O) emission in an analytically-tractable manner. The values for the dimensionless temperatures $\Theta_i$ and weights $f_i$ are given in Table~\ref{tab:radiation_params}, where adopted values represent a maximum plausible EBL model to consider the greatest attenuating effect on our results. Here, the redshift weighting evolutionary function is given by \begin{equation}
  \mathcal{K}(z) = \begin{cases}
    (1+z)^{3.1} \ , \hspace{0.9cm} z<z_{\rm cut}\\
    (1+z_{\rm cut})^{3.1} \ , \hspace{0.5cm} z\geq z_{\rm cut}
    \end{cases}
    \label{eq:kevo}
\end{equation}
up to $z=3$, where $z_{\rm cut} = 1.4$, which follows the `baseline' EBL redshift evolution model of \cite{Aloisio2013APh}. 

\subsection{Nuclei path lengths}
\label{subsec:nuclei_path_lengths}

\begin{figure}
    \centering
    \includegraphics[width=0.8\columnwidth]{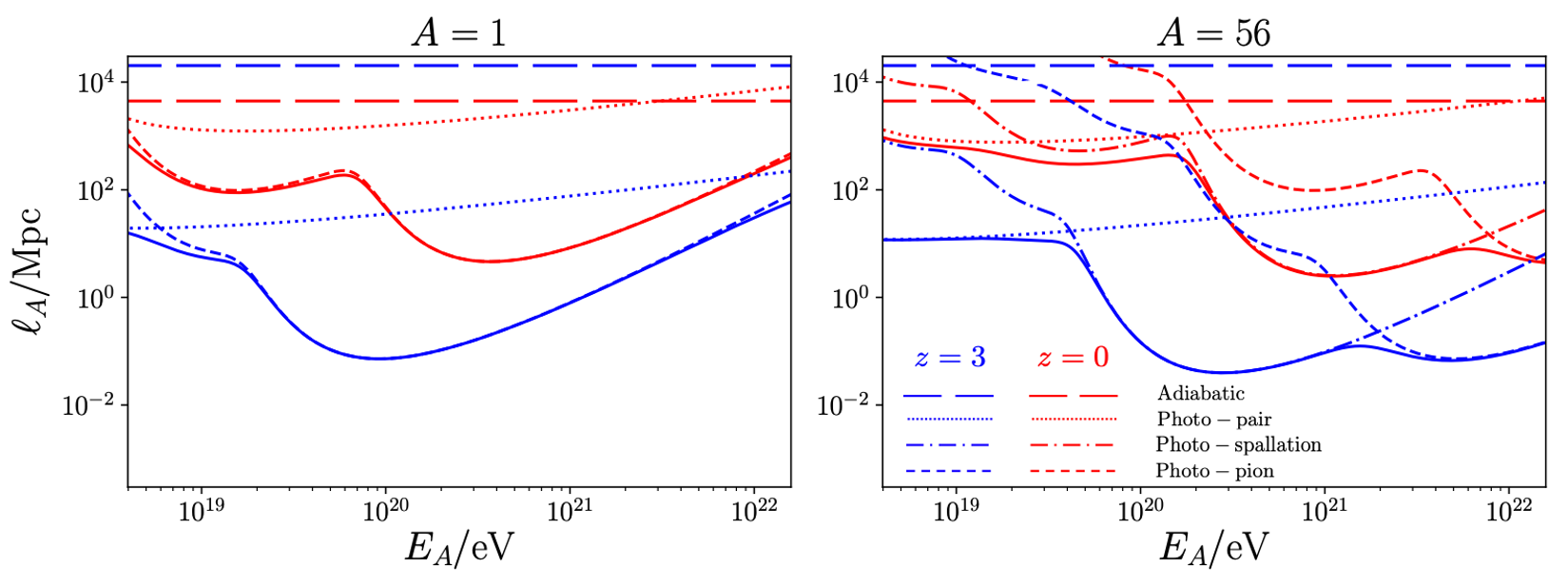}   
        \caption{The characteristic lengths of energy-loss   
  for nuclei traversing through background (CMB and EBL) cosmological radiation fields for various cooling and absorption processes (as labeled). Red lines denote distances computed at $z=0$, while blue lines denote distances computed at $z=3$. The left panel shows the case for $^{1}$H, while the right panel shows the energy-loss lengths for $^{56}$Fe. As $^{1}$H is comprised only of 1 nucleon, it cannot be further degraded by photo-spallation.}
  \label{fig:path_lengths}
\end{figure}

\noindent
The relative importance of the cooling and interaction processes of UHE CRs described in section~\ref{sec:section_21} in the EBL and CMB can be characterized in terms of an effective energy-loss path length, $\ell_A$.
This is the characteristic distance over which a CR nucleus of energy $E_A$ would lose its energy due to its interactions, and can be estimated from the CR cooling rate $b$ for a given process using $\ell_A = c\;\![|b|/\epsilon_{A}]^{-1}$.
  In the case of stochastic absorption processes, $\ell_A$ may be approximated as a cooling rate for the purposes of comparison, i.e. 
  ${b^{A \pi}}_{\! A} \approx -\epsilon_A{\Lambda^{A \pi}}_{\! A}$ 
    for pion-production, 
    or as ${b^{\rm sp}}_{\!A} \approx -\epsilon_A{\Lambda^{\rm sp}}_{\!A}$ for photo-spallation.\footnote{Strictly, this is only appropriate for the purposes of rough comparison between different interactions. Absorption interactions are stochastic in nature, and are not truly continuous cooling processes. This would introduce statistical broadening to the absorption locations, which can change the mean path length by a factor of a few, and is particularly severe at high energies. This is properly accounted for in our later calculations.}
  In {Figure}~\ref{fig:path_lengths}, 
   we compare the effective loss lengths for UHE CR nuclei ($A=1$ for protons and $A=56$ for $^{56}$Fe nuclei) 
   in intergalactic conditions at both $z=0$ and $z=3$. This shows that the more intense background radiation fields at $z=3$ compared to $z=0$ reduce photo-pion and photo-spallation path lengths, however adiabatic losses are noticeably weaker at $z=3$ due to the accelerated rate of cosmic expansion in the later Universe. Radiative losses are inconsequential in all cases.

\section{CR source models}
\label{sec:section_3}

\begin{figure}
    \centering
    \includegraphics[width=0.8\columnwidth]{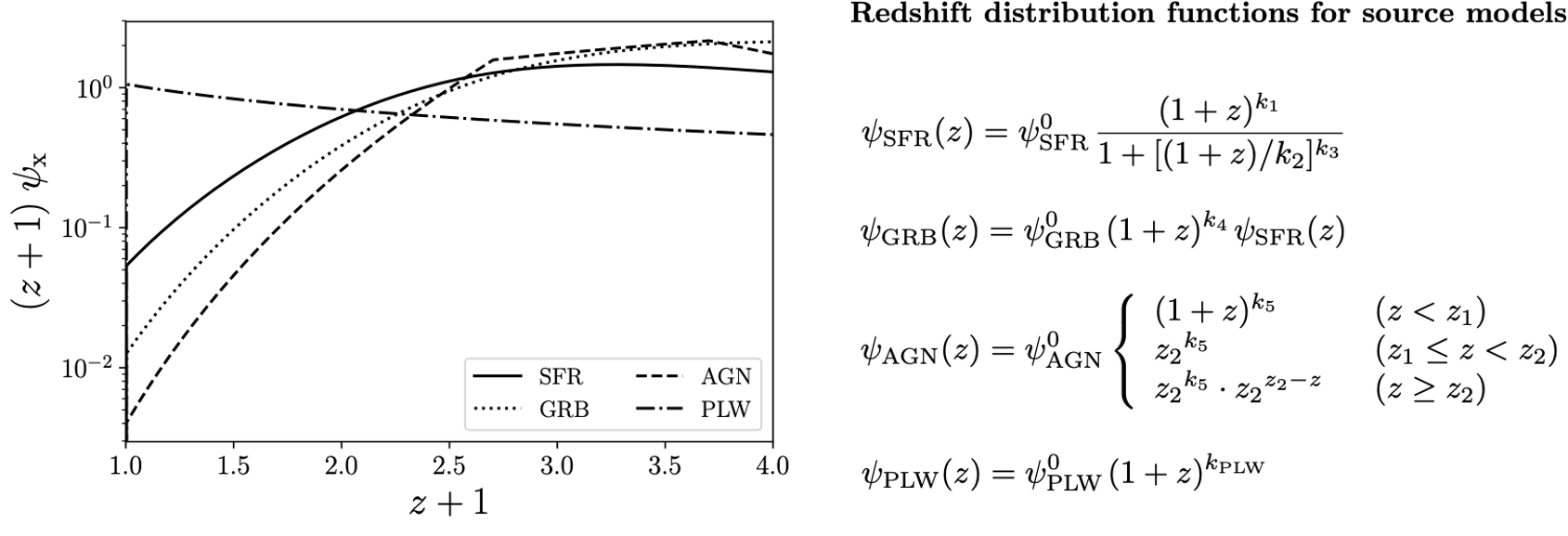}   
        \caption{Redshift distributions of 
        the four source models,  
        SFR, GRB, AGN and PLW.   
        Their corresponding normalized redshift evolution functions are indicated where, for the \textbf{SFR model} $k_1 = 2.7$, $k_2 = 2.9$ and $k_3 = 5.6$, for the \textbf{GRB model} $k_4 = 1.4$, for the \textbf{AGN model} $k_5 = 5.0$, $z_1=1.7$, $z_2=2.7$, and for \textbf{PLW model} $k_{\rm PLW} = -1.6$ (see \cite{Batista2019JCAP} for discussion of these parameter choices). $\psi^0_{\rm x}$ values are given in Table~\ref{tab:parameters}.}
  \label{fig:redshift_distributions}
\end{figure}

\noindent
The UHE CR injection rate by source populations may be written as:
\begin{align}
\label{eq:astrophysical_injection}
  {\mathcal{Q}^{\rm a}}_{\! A}(z,\epsilon_A, A)
    =\mathcal{C}_0(1+z)^{-3} \psi_{\rm x}(z)\;\!\frac{\mathrm{d}}
    {\mathrm{d}\epsilon_A}
    \left(\frac{\mathrm{d}\;\!{n}^{\star}}{\mathrm{d}\;\!t} \right) \ , 
\end{align}
where ${\mathcal{Q}^{\rm a}}_{\!A}$ and ${Q^{\rm a}}_{\!A}$ 
  are the comoving and physical source terms, respectively. $\psi_{\rm x}(z)$ 
  is the evolutionary function 
  describing the distribution of a CR source population with respect to redshift $z$, up to $z_{\rm max} = 3$ (the specific model is denoted by ${\rm x}$).  
  We consider four possible source models, as described in the annotation in  Figure~\ref{fig:redshift_distributions}. 
  The normalization $\mathcal{C}_0$ is defined according to the UHE CR luminosity density,
  \begin{align}
\label{eq:overall_normalisation}
  \rho_{\rm CR} 
    &= \mathcal{C}_0\sum_{A}
   \int_{\epsilon_{\mathrm{min}}}^{\epsilon_{\mathrm{max}}} 
 \mathrm{d}\;\!\epsilon_A \  
  \epsilon_A\;\! \psi_{\rm x}(z)\;\!  
  \frac{\mathrm{d}}{\mathrm{d}\;\!\epsilon_A}
  \left(\frac{\mathrm{d}\;\!{n}^{\star}}{\mathrm{d}\;\!t} \right)\;\!
  \biggr \vert_{z=0}  \ , 
\end{align}
which we treat as a model parameter, with values given by Table~\ref{tab:parameters} for each considered source population. Here, 
  $\mathrm{d}/\mathrm{d}{\epsilon_A}\;\!(\mathrm{d}n^{\star}/ \mathrm{d} t) = \mathrm{d}/\mathrm{d}{t}\;\!(\mathrm{d}n^{\star}/ {\mathrm{d}\epsilon_A})$ is the volumetric spectral injection rate of CRs by a given source population, with a parametrised injection spectrum
  \begin{align}%
\label{eq:normalisation_injection}
\frac{\mathrm{d}\;\!n^{\star}}{\mathrm{d}\;\!\epsilon_A} 
 & \propto f_{A} \left(\frac{\epsilon_A}{\epsilon_{\rm min}}\right)^{-\alpha} 
\left\{ 
\begin{array}{ll}
1 & ( x<1)  \\
\exp \left(1-x\right) & (x\geq 1) \
\end{array}\right.   \   
\end{align} 
\cite[e.g.][]{Batista2019JCAP}, 
where $x = {\epsilon_{\rm A} m_{\rm e} c^2}/{Z_A R_{\max }}$,
   $R_{\rm max}(V)$ is the rigidity, and 
  the energy limits are   
  $\epsilon_{\mathrm{min}} m_{\rm e} c^2 = 3.98 \times 10^{18}\;\!{\rm eV}$ and 
   $\epsilon_{\mathrm{max}} m_{\rm e} c^2 = 3.16\times 10^{20}\;\!{\rm eV}$. 
The values of the spectral index $\alpha$ 
 and the rigidity are shown in  
{Table}~\ref{tab:parameters}. 
The spectral composition adopts fitted values obtained by~\cite{Batista2019JCAP}, where the full range of injected species are represented by abundances of $^1$H, $^4$He, $^{14}$N, $^{28}$Si and $^{56}$Fe.

\begin{table}
\centering
\resizebox{0.8\textwidth}{!}{%
\begin{tabular}{lccccc}
\hline \hline
{\bf Model} & {\bf Normalization $\psi_{\rm x}^{0}$} & {\bf Spectral index $\alpha$} & {\bf ${\rm log}(R_{\rm max}/V)$} & {\bf $\rho_{\rm CR}$/$10^{48}$ erg Mpc$^{-3}$ yr$^{-1}$}\\ \hline
    (1) SFR & $\psi_{\rm SFR}^{0}= 0.054$ & $-1.3$ & 18.2 & 0.5 \\
    (2) GRB & $\psi_{\rm GRB}^{0}= 0.013$ & $-1.5$ & 18.2 & 2.0 \\
    (3) AGN & $\psi_{\rm AGN}^{0}= 0.0041$ & $-1.0$ & 18.2 & 0.04 \\
    (4) PLW & $\psi_{\rm PLW}^{0}= 1.1$ & $+1.0$ & 18.7 & 15.0 \\
    \hline \hline
\end{tabular}}
\caption{Summary of parameter choices adopted for each of the four redshift source distribution models.} 
\label{tab:parameters} 
\end{table}

\section{Results}
\label{sec:section_4}

\noindent
The CR spectrum, expected to be observed at $z=0$, is computed by numerically solving the transport equation under a quasi-steady 
  condition for each of the four source models: 
  \begin{align}
\label{eq:transport_equation_general}
\frac{\partial\;\! {n_A}}{\partial z} 
  =  \frac{1}{c}
    \left[\frac{\partial}{\partial \epsilon_A} 
    \left(b_A n_A \right)+ {Q}_A -  \Lambda_A n_A \right]
    \frac{{\rm d} s}{{\rm d} z} \ .
\end{align}
 The numerical integration proceeds    
   from $z_{\rm max}=3$ to $z_{\rm min}=0$. 
 Here, ${{\rm d}s}/{{\rm d}z} = {c~\mathcal{E}(z)}/{H_0\;\!(1+z)}$, primary CR injection as well as secondary nuclear production arising from interactions was encoded into the source term, where $Q_A = {Q^{\rm sp}}_{\! A}(z, \epsilon_A) + {Q}^{\rm a}_{\! A}(z,\epsilon_A)$, for ${Q}^{\rm a}_{\! A} = {\mathcal{Q}^{\rm a}}_{\! A}(z,\epsilon_A, A) (1+z)^{3}$, where ${\mathcal{Q}^{\rm a}}_{\! A}$ is given by equation~\ref{eq:astrophysical_injection} and ${Q^{\rm sp}}_{\! A}$ is given by equation~\ref{eq:source_secondary}. 
   
\begin{figure*}
    \centering
    \includegraphics[width=0.8\textwidth]{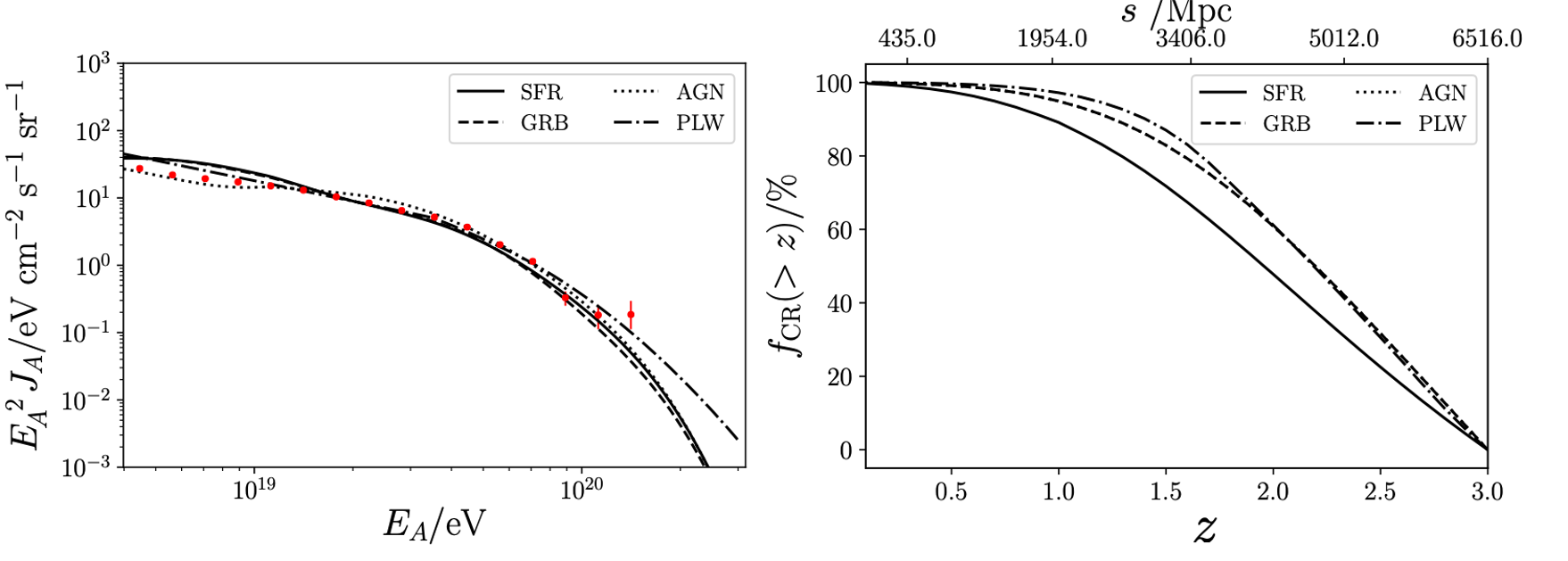}   
        \caption{\textbf{Left}: Flux spectra of CRs for the source models,  
          SFR, GRB, AGN and PLW propagated to $z=0$ for all species. The data (discrete red data points with error bars) 
       obtained by the Pierre Auger Observatory 
       \citep{Valerio2019APoS} 
       are shown for comparison. 
        \textbf{Right}: The fraction of UHE CRs (between energies of $3.98 \times 10^{18}\;\!{\rm eV}$ to 
   $3.16\times 10^{20}\;\!{\rm eV}$)
          that originate from a  redshift higher than $z$,   
          $f_{\rm CR}(>z)$, for the four source models. 
        The bottom abscissa shows the redshift $z$, 
          and the top abscissa shows the corresponding comoving distances.
}
\label{fig:cr_distances}
\end{figure*}

   Figure~\ref{fig:cr_distances} shows the UHE CR spectrum at $z=0$ in the case of the four source population models in the left panel. Data 
       obtained by the Pierre Auger Observatory 
       \citep{Valerio2019APoS} 
       are included for comparison (for more detailed comparisons with data, see~\cite{Owen2021arXiv210712607O}). In the right panel of Figure~\ref{fig:cr_distances}, the fractional contribution from sources  
  above a given redshift, $f_{\rm CR}(>z)$,
  for the four source models is shown. This shows, in all cases, a large faction of UHE CR flux observed at Earth can not be attributed to local sources, regardless of the uncertainties in existing models of the EBL at high redshift or CR composition. Moreover, the curves for the GRB and AGN models
  are practically indistinguishable. This is because the curve is a manifestation of where the redshift locations of the underlying source population make the most significant contribution to the UHE CR flux. The dominant redshift locations 
  of the GRB and AGN CR sources 
  are very similar to one another (see Figure~\ref{fig:redshift_distributions}) which, together with the  
  small numbers of GRBs and AGN 
  below $(z+1) \sim 2.5$ 
  leads to their almost identical $f_{\rm CR}(>z)$ curves.  

\section{Conclusions}
\label{sec:section_5}

\noindent
Our calculations show that 
distant UHE CR sources at redshifts as high as $z\sim (2-3)$ 
  contribute substantially 
  to UHE CRs detected on Earth, naturally leading to the formation of a strong, isotropic background in the UHE CR flux. 
Moreover, we find that most of the UHE CRs from these distant sources  
  are primary particles,  
  despite the large cosmological distances they have traversed, with their spectra and composition at $z=0$ being almost indistinguishable from that injected by the source population, even when fully accounting for photo-spallation of nuclei along CR propagation paths.  
  We find our results to be robust, and not strongly dependent on CR composition, source redshift distribution (if reasonable) or EBL intensity up to redshifts as high as $z\sim 3$. 
 
\acknowledgments

\noindent
ERO is supported by the Center for Informatics and Computation in Astronomy (CICA) at National Tsing Hua University (NTHU), funded by the Ministry of Education (MoE) of Taiwan. YXJY is supported by a NTHU International Student Scholarship, and by a grant from the Ministry of Science and Technology (MoST) of Taiwan, 109-2628-M-007-005-RSP. This work used high-performance computing facilities operated by CICA at NTHU, funded by MoE and MoST.

\bibliographystyle{ICRC}
\bibliography{references}

\end{document}